\documentclass{iaus}
\usepackage{graphicx}
\usepackage{natbib}

\title[The Evolution of the bTFR over the past 6 Gyr] %% give here short title %%
{The Evolution of the Baryonic Tully-Fisher Relation over the past 6 Gyr}

\author[M. Puech et al.]   %% give here short author list %%
{M. Puech$^1$, F. Hammer$^1$, H. Flores$^1$, R. Delgado-Serrano$^{1,2}$, M. Rodrigues$^{1,4,5}$, \and Y. Yang$^{1,5}$
}

\affiliation{$^1$GEPI, Observatoire de Paris, CNRS, University Paris Diderot; 5 Place Jules Janssen, 92190 Meudon, France \\ email: {\tt mathieu.puech@obspm.fr} \\[\affilskip]
$^2$Panama Observatory, Technological University of Panama, 0819-07289 Panama, Rep. of Panama\\
$^4$CENTRA, Instituto Superior Tecnico, Av. Rovisco Pais 1049-001 Lisboa, Protugal \\
$^5$ European Southern Observatory, Casilla 19001, Santiago 19, Chile\\
$^6$ National Astronomical Observatories, Chinese Academy of Sciences, 20A Datun Road, Chaoyang District, Beijing 100012, China\\
}

\pubyear{2011}
\volume{277}  %% insert here IAU Symposium No.
\pagerange{119--126}
\jname{Tracing the Ancestry of Galaxies on the Land of our Ancestors}
\editors{C. Carignan, K.C. Freeman \& F. Combes, eds.}
\begin{document}

\maketitle

\begin{abstract}
Scaling relations are salient ingredients of galaxy evolution and
formation models. I summarize results from the IMAGES survey, which
combines spatially-resolved kinematics from FLAMES/GIRAFFE with
imaging from HST/ACS and other facilities. Specifically, I will focus
on the evolution of the stellar mass and baryonic Tully-Fisher
Relations (TFR) from z=0.6 down to z=0. We found a significant
evolution in zero point and scatter of the stellar mass TFR compared
to the local Universe. Combined with gas fractions derived by
inverting the Schmidt-Kennicutt relation, we derived for the first
time a baryonic TFR at high redshift. Conversely to the stellar mass
TFR, the baryonic relation does not appear to evolve in zero point,
which suggests that most of the reservoir of gas converted into stars
over the past 6 Gyr was already gravitationally bound to galaxies at
z=0.6. 
\end{abstract}

\firstsection % if your document starts with a section,
\section{Introduction}
The distant Tully-Fisher Relation (TFR) has been extensively studied
over the past two decades. One of its most intriguing property is that
it shows a significantly larger scatter at high redshift compared to
the local Universe. This effect is seen both in B-band and in K-band
(e.g., \citealt{bohm07}, \citealt{conselice05}), which indicates that
it is not related to dust or stellar population effects.

Integral Field Spectroscopy (IFS) now allows us to study the
spatially-resolved kinematics of distant galaxies. First studies of
the distant TFR using IFS revealed a correlation between the scatter
of the K-band TFR at z$\sim$0.6 and the dynamical state of galaxies,
meaning that the larger the scatter in the TFR, the larger the
dynamical disturbances observed in the galaxy kinematics
\citep{flores06}. It was shown that once reduced to rotating disks,
the z$\sim$0.6 K-band TFR has the same scatter that the local
relation, which indicates that the scatter in the distant TFR is
indeed entirely due to kinematic disturbances, strong enough to
destroy rotation on large spatial scales. This led \cite{puech08} to
suggest that major mergers are the physical driver for such kinematic
disturbances in a large fraction of z$\sim$0.6 galaxies (see also
\citealt{neichel08} from a morphological point of view).

Interestingly, a new kinematic tracer called $S$ was proposed by
\cite{weiner06}. This tracer combines both the rotation velocity and
the velocity dispersion, which makes it a tracer of the total
kinematic energy. \cite{kassin07} showed that, when applied to the
TFR, one gets a new relation (hereafter $S$-TFR), on which the scatter
is significantly reduced compared to the ``regular'' TFR (i.e., when
the rotation velocity only is considered). \cite{covington09} used
hydrodynamical simulations to suggest that such an effect on the
scatter could result from shocks in the gaseous phase during galaxy
major mergers. These shocks would lead to a transfer of energy from
bulk motions (i.e., large-scale rotation) to random motions, occuring
at smaller spatial scales. This transfer of energy translates into a
drop in the rotation velocity as a function of time while the velocity
dispersion increases, which results in $S$ staying constant at first
order. This would explain why the $S$-TFR shows a much reduced scatter
compared to the regular TFR.

\section{The IMAGES stellar-mass (smTFR) and $S$-TFR}
We studied both the stellar-mass (smTFR) and $S$-TFR in the IMAGES
sample. The IMAGES sample is a representative sample of 63 M* emission
line galaxies selected in four different fields of view (i.e., CFRS03,
CFRS22, HDFS, and CDFS, see \citealt{yang08}). All galaxies were
observed using the FLAMES/GIRAFFE IFUs, which allowed us to map their
kinematics at relatively large spatial scales ($\sim$3.5 kpc/pix at
z$\sim$0.6). We derived the rotation velocities using Monte-Carlo
simulations, taking carefully into account beam smearing effects
\citep{puech08}. Stellar masses were obtained from (dust and
inclination corrected) K-band luminosities and B-V colors using the
\cite{bell03} simplified prescriptions. We assumed a diet Salpeter
IMF, which, in combination to the \cite{bell03} prescriptions for
stellar masses, has the interesting property of maximizing the stellar
mass estimates, meaning that any other combination of stellar
population model and IMF would lead to smaller masses
\citep{hammer09}. This means that any shift obtained with this
combination between the distant and local relation would be larger
with any other combination.

The derived smTFR at z$\sim$0.6 is shown on the left panel of Fig.~1.
It confirms the results of \cite{puech08}, that the entire scatter of
the distant relation is due to dynamically non-relaxed galaxies. Once
reduced to rotating disks, the smTFR is shifted downward by
0.34$^{0.21}_{-0.06}$ dex (where the quoted errobars represent
systematic uncertainties) compared to the local relation, which
suggests that intermediate-mass, emission line rotating disks grew up
by a factor $\sim$2 in stellar mass over the past 6 Gyr. This shift is
robust against both random and systematic uncertainties (see
\citealt{puech10} for details).

Similar results were obtained at z$\sim$2.2 in the SINS survey:
\cite{cresci09} found a z$\sim$2.2 relation shifted downward by 0.41
dex compared to the location relation. At first sight, it might
however be surprising to find a similar shift at z$\sim$0.6 and
z$\sim$2 compared to the local relation. We showed in \cite{puech10}
that this is linked to the local sample used to derive the shift in
zero point. Several widely used local comparison samples were indeed
compared in \cite{hammer07}, where it was shown that the
\cite{verheijen01} Ursa Major sample is biased towards low-mass
galaxies, when compared to the local velocity function. Other samples,
e.g., the \cite{courteau97} sample drawn from the UGC catalog or the
\cite{pizagno07} drawn from the SDSS, appear to more appropriate when
one compares representative samples at different redshifts. In the
IMAGES survey, we used the SDSS sample as a reference, while the SINS
survey relied on the \cite{bell01} local smTFR, which was drawn from
the \cite{verheijen01} sample. If one re-derives the shift between the
z$\sim$2.2 smTFR and the SDSS smTFR relation, the resulting offset
increases from $\sim$0.4 to $\sim$0.6 dex, i.e., significantly larger
than the shift of 0.36 dex found between the z$\sim$0.6 smTFR and the
local relation. When comparing the TFR between different redhifts, it
is therefore very important to adopt representative samples, or at
least samples which share the same properties in terms of mass and/or
velocity distribution.

We also derived the $S$-TFR in the IMAGES sample (see middle panel of
Fig.~1), and confirms the results of \cite{kassin07}, i.e., the
scatter is reduced by a factor $\sim$2 compared to the regular smTFR.
We also found observational evidence for the transfer of energy
suggest by \cite{covington09}. One of the object in the IMAGES sample
was indeed found to be caracteristic of the shrinking effect in the
scatter due to the use of $S$ (see Fig.~1): it was modeled in detail
using hydrodynamical simulations by \cite{peirani09}, who found that
the best model for this object corresponds to a 3:1 major merger.

\section{The IMAGES baryonic TFR}
We estimate gas masses in the IMAGES sample by inverting the
Schmidt-Kennicutt law between star formation rate and gas mass
densities \cite{kennicutt89}. Total SFRs were estimated by summing
contributions from the UV and the IR, while total gas radii were
estimated using GIRAFFE-IFU [OII] maps carefully deconvolved from beam
smearing effects (see \citealt{puech10}). Interestingly, the ionized
gaz radius is found to be on average $\sim$30\% more extended than the
total UV radius, which suggests that another physical process than
radiation from OB stars is ionizing gaz at large radii. We found
observational evidence that shocks during major mergers could be this
alternative driver \citep{puech09}. Such a process could also be
responsible for the increased velocity dispersion observed in
z$\sim$0.6 disks compared to local ones \citep{puech07, epinat10}, as
also observed in higher-redshif disks \citep{forster09}.

Galaxies at z$\sim$0.6 show a median gas fraction of $\sim$31$\pm$1\%.
The same mean gas fraction was estimated independantly by
\cite{rodrigues08} relying on the evolution of the mass-metallicity
relation combined with a closed-box chemical model. This suggests that
systematic effects are probably limited and gives us some confidence
in the estimated mass of gas in z$\sim$0.6 galaxies.

The resulting baryonic TFR (bTFR) is shown on the right panel of
Fig.~1. This is the first time that the bTFR is derived at high
redshifts. On overall, it shows the same structure that the smTFR.
However, conversely to the smTFR that is shifted downward by a factor
$\sim$2 in stellar mass, the bTFR does not show any evidence for
evolution in zero point compared to the local relation. This indicates
that the reservoir of gas that has been converted into stars over the
past 6 Gyr was already gravitationnally bound to disks at z$\sim$0.6.
As a consequence, there is no need for external gas accretion, in
qualitative agreement with cosmological simulations which predict that
cold flows vanish at z$\leq$1.5 in massive halos where such galaxies
inhabit in \citep{keres09,puech10b}. This does not mean that there is
no gas accrection in these galaxies since within systematic
uncertainties, there is room for cold gas accretion at a level of
$\sim$30\% of the local baryonic mass, but there is presently no need
for such an accretation, at least from a dynamical point of view.

\begin{figure}[!ht]
\begin{center}
\includegraphics[scale = 0.22]{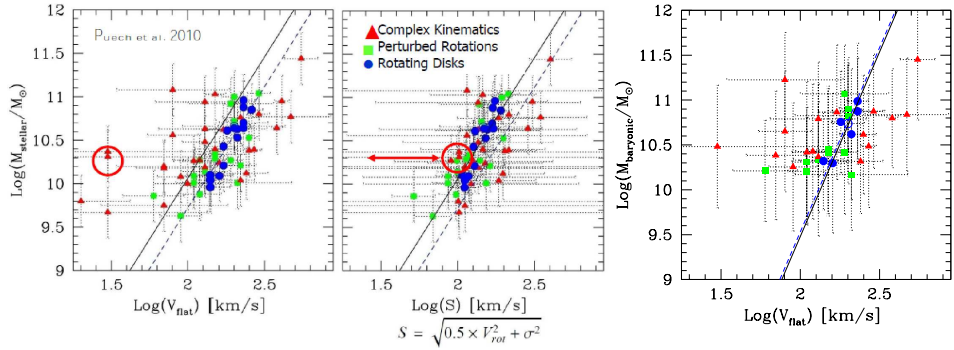}
\end{center}
\caption{\emph{Left}: Evolution of the stellar-mass TFR between
  z$\sim$0.6 (dashed line) and z=0 (solid line). Distant rotating
  disks are shown as dots, perturbed rotators as squares, and galaxies
  with complex kinematics as triangles. \emph{Middle}: Same relation
  but using the kinematic tracer S, which combines rotation velocity
  with velocity dispersion (see \citealt{kassin07}). The solid and
  dashed lines are the same that in the left panel. \emph{Right}:
  Evolution of the baryonic TFR in the CDFS sub-sample. The solid line
  is the local bTFR, while the dashed line is a fit to the distant
  relation. See \cite{puech10} for details. The object encircled in
  red in the two leftest panels correspond to a major merger modeled
  in detail by \cite{peirani09} using hydrodynamical simulations (see
  text).}
\end{figure}

%\begin{discussion}

%\discuss{Maraston}{question about stellar mass}

%\end{discussion}


\begin{thebibliography}{}
\bibitem[Bell \& de Jong(2001)]{bell01} Bell, E.F., \& de Jong, R.S.
    2001, \textit{ApJ}, 550, 212
\bibitem[Bell et al.(2003)]{bell03} Bell, E.~F., McIntosh, D.~H.,
  Katz, N., \& Weinberg, M.~D.\ 2003, \textit{ApJS}, 149, 289
  \bibitem[B\"ohm \& Ziegler(2007)]{bohm07} B\"ohm, A., \& Ziegler,
  B.L. 2007, \textit{ApJ}, 668, 846
\bibitem[Conselice et al.(2005)]{conselice05} Conselice, C.~J.,
Bundy, K., Ellis, R.~S., Brichmann, J., Vogt, N.~P., \& Phillips,
A.~C.\ 2005, \textit{ApJ}, 628, 160
\bibitem[Courteau(1997)]{courteau97} Courteau, S. 1997, \textit{AJ}, 114,
  2402
\bibitem[Covington et al.(2009)]{covington09} Covington, M.D., Kassin,
  S.A., Dutton, A.A., et al. 2009, \textit{ApJ}, 710, 279
\bibitem[Cresci et al.(2009)]{cresci09} Cresci, G., Hicks, E.K.S.,
    Genzel, R., et al. 2009, \textit{ApJ}, 697, 115
  \bibitem[Epinat et al.(2010)]{epinat10} Epinat, B., Amram, P.,
  Balkowski, C., \& Marcelin, M. \textit{MNRAS}, 401, 2113
\bibitem[Flores et al.(2006)]{flores06} Flores, H., Hammer, F.,
    Puech, M., Amram, P., \& Balkowski, C.\ 2006, \textit{A\&A}, 455, 107
\bibitem[F{\"o}rster Schreiber et al.(2009)]{forster09} Forster
    Schreiber, N.~M., et al.\ 2009,\textit{ApJ}, 706, 1364
\bibitem[Hammer et al.(2007)]{hammer07} Hammer, F., Puech, M.,
    Chemin, L., Flores, H., \& Lehnert, M.\ 2007, \textit{ApJ}, 662, 322
\bibitem[Hammer et al.(2009)]{hammer09} Hammer, F., Flores, H., Puech,
  M., et al. 2009, A\&A, 507, 1313
\bibitem[Kassin et al.(2007)]{kassin07} Kassin, S.A., Weiner, B.J.,
  Faber, S.M., et al. 2007, \textit{ApJ}, 660, 35
\bibitem[Kennicutt(1989)]{kennicutt89} Kennicutt, R.\ 1989, \textit{ApJ},
  344, 685
\bibitem[Kere{\v s} et al.(2009)]{keres09} Kere{\v s}, D., Katz, N.,
  Fardal, M., Dav{\'e}, R., \& Weinberg, D.~H.\ 2009, \textit{MNRAS},
  395, 160
\bibitem[Neichel et al.(2008)]{neichel08} Neichel, B., et al. 2008,
  \textit{A\&A}, 484, 159
\bibitem[Peirani et al.(2009)]{peirani09} Peirani, S., Hammer, F.,
    Flores, H., et al. 2009, \textit{A\&A}, 496, 51
\bibitem[Pizagno et al.(2007)]{pizagno07} Pizagno, J., Prada, F.,
  Weinberg, D.H., et al. 2007, \textit{AJ}, 134, 945
\bibitem[Puech et al.(2007)]{puech07} Puech, M., Hammer, F.,
  Lehnert, M.~D., \& Flores, H.\ 2007a, \textit{A\&A}, 466, 83
\bibitem[Puech et al.(2008)]{puech08} Puech, M., Flores, H., Hammer,
  F., et al. 2008, \textit{A\&A}, 484, 173
\bibitem[Puech et al.(2009)]{puech09} Puech, M., Hammer, F.,
  Flores, H., Neichel, B., Yang, Y. 2009, \textit{A\&A}, 493, 899
\bibitem[Puech et al.(2010)]{puech10} Puech, M., Hammer, F., Flores, H., et al. 2010, \textit{A\&A}, 510, 68
\bibitem[Puech(2010)]{puech10b} Puech, M. 2010, \textit{MNRAS}, 406, 535
\bibitem[Rodrigues et al.(2008)]{rodrigues08} Rodrigues, M., Hammer,
    F., Flores, H., et al. 2008, \textit{A\&A}, 492, 371
\bibitem[Verheijen(2001)]{verheijen01} Verheijen, M.A.W. 2001, \textit{ApJ},
    563, 694
\bibitem[Weiner et al.(2006)]{weiner06} Weiner, B.~J., et al.\ 2006,
  \textit{ApJ}, 653, 1027
\bibitem[Yang et al.(2008)]{yang08} Yang, Y., Flores, H., Hammer,
    F., et al. 2008, \textit{A\&A}, 477, 789
\end{thebibliography}
\end{document}